\documentclass[a4paper, 11pt]{article}
\RequirePackage[margin=15mm]{geometry}
\RequirePackage{fancyhdr}

\newcommand{\keywords}[1]{
\medskip
Keywords: \textit{#1}
}

\newenvironment{affiliations}{
\medskip
}

\renewenvironment{abstract}{
\small
\medskip\medskip
}

\usepackage{datetime}
\usepackage[utf8]{inputenc}
\usepackage[noenc]{tipa}
\usepackage{amssymb,amsfonts,textcomp,amsmath}
\usepackage[T3,T1]{fontenc} 
\usepackage{csquotes}
\usepackage{hyperref} 
\usepackage{url}
\usepackage{amsmath}
\usepackage[english]{babel}
\usepackage{xr}  
\usepackage{caption} 

\hypersetup{colorlinks=true, linkcolor=red, citecolor=red, filecolor=blue,
  urlcolor=blue, pdftitle=, pdfauthor=Georges Sitja, pdfsubject=, pdfkeywords=}

\begin{document}

\pagestyle{fancy}
\fancyhead{} 
\fancyfoot{} 
\fancyfoot[R]{G.S. ~-~ \the\day-\the\month-\the\year ~-~\xxivtime}
\fancyhead[R]{\thepage}
\fancyhead[L]{\leftmark \\ \rightmark} 
\footskip = 40pt
\textheight = 730pt
\title{Modifications of statistics under dimer diffusion}

\date{\vspace{-5ex}} 

\maketitle



\begin{affiliations}
Georges Sitja\\
CNRS, UMR 7325, Aix-Marseille University, Cinam, Campus Luminy, Case 913,\\
F-13288 Marseille 09, France.
\end{affiliations}


\color{black}

\begin{abstract}

The diffusion, ``explosion'' and ``evaporation'' of dimers and the subsequent coalescence are treated in a formal way by identifying and solving the differential equations deduced from the respective behaviors of dimers in the different cases. This study leads to analytic formulas allowing to calculate, in a simple and fast way, the size statistics obtained after the coalescence of the dimers or their constituents once the dimers have completely disappeared. These formulas are of capital interest to characterize systems in which the dimers initially present disappear.

\end{abstract}

\section{Introduction}

On December 29th, 1959, Richard Feynman gave a speech \cite{Feynman_1960} on the still unexplored
properties of infinitely small objects.  He specifies that by ``small'' he does not mean objects
like millimeter-sized electric motors which already exist in 1960, but of structures whose
characteristic sizes are of the order of few nanometers. Although the main part of his speech
speaks about miniaturization, he foresees the quite singular properties of the nanometric objects,
making appear quantum effects. He also makes an allusion to the problem that could constitute the
diffusion for the stability of such objects.

Nowadays, the study of the properties of nanometric clusters of atoms has become common in various
fields, such as optics\cite{ISI:A1983QP29700020}\cite{ISI:000166122000006},
magnetism\cite{ISI:000232992100001}, and heterogeneous catalysis\cite{ISI:A1984ABP3500010}.
The search for new properties has led to the study of smaller and smaller metal particles,
approaching, or even reaching the ultimate size of the individual atom\cite{ISI:000413190800003}.
The use of nano-structured supports makes that, nowadays, many scientific studies are conducted
on cluster networks at critical sizes where diffusion is likely to occur\cite{ISI:000440505000051}%
\cite{ISI:000488835000007}\cite{ISI:000458704800047}\cite{ISI:000292999100017}%
\cite{ISI:000433404300004}. When depositing atoms on a nano-structured surface, the Poisson
distribution\cite{ISI:000274979000024} can be achieved by choosing proper growth parameters
(substrate temperature, deposition rate).
This is interesting, not only because of the satisfaction provided by the knowledge of the sample state, but also because it's extremely useful to deduce the properties of the clusters. Indeed, by measuring the activity of an assembly of particles and by varying the quantities of deposited atoms, the knowledge of the Poisson distribution allows us to deconvolute the signal to obtain the responses of the clusters as a function of their size. However, if during the experiments, because of the temperature or the chemical environment, diffusion occurs, the initial well-characterized system disappears leadingand is replaced by a new and inoperable system.
The experimental conditions can be sufficient
to initiate the diffusion of monomers\cite{ISI:000488835000007}, dimers\cite{ISI:000242219900018},
but also larger clusters\cite{ISI:000458704800047}. At present, the most common technique to
determine the size statistics of very small clusters is to use an STM (Scanning Tunneling Microscope)
which provides safely the ratio between occupied and unoccupied sites, i.e. the occupancy rate.

In the litterature, the most advanced formula to determine the occupancy rate after a deposition assuming that monomers
and dimers have diffused is given in an article by Liu et al.\cite{ISI:000366958900001}. This formula gives the occupancy rate $R_{occ}$ assuming the initial distribution is known, but does not tell us anything about the final size distribution.

\begin{equation}
  \label{formule_bidon}
  R_{occ} = 1 - P_1 - P_2
\end{equation}

Where $P_n$ are the known initial probabilities of having a site occupied by a cluster of n atoms. However, this formula is eminently false since it does not take into account the formation of new particles of size greater than 2 during the coalescence of dimers with monomers or dimers with dimers. One can easily convince oneself that by thinking about a deposit where only monomers and dimers are present, the occupancy rate will be 0 after the diffusion. It is strange that such kind of bogus formula can be pulled out of a hat. It is also amazing that this bogus formula is used to question the experimental measurements of another research team.

\section{Motivations}

This study is motivated by the following concrete problem (which will serve as an experimental reference when it will not be specified otherwise): One makes a deposit of atoms (for example, by condensation of a flow of atoms coming from an evaporator) on a surface having nucleation centers distributed on a network. In this case, the environment of each nucleation center is identical and the probability of capture of an atom diffusing on the surface is the same whatever the nucleation center considered. The first question that can be asked is "What is the size distribution of the clusters formed as a function of the mean number of atoms per site deposited? In fact, the answer to this question is known: If an average of x atoms per nucleation site have been deposited, the probability of having a cluster of n atoms follows the Poisson law\cite{poisson}:

\begin{equation}
  P_n(x)=\frac{1}{n!}x^ne^{-x}
\end{equation}

The second question is: What is the new size distribution if during the experiments the monomers or the dimers start to diffuse and disappear to leave only clusters of size greater than 2 atoms. The diffusion of monomers is already addressed in an exact way\cite{sitja2021statistics}, and I will focus here on the dimer diffusion.

\section{How to get rid of dimers?}

Before starting the calculations, it is good to think about the different scenarios that eliminate dimers from the surface. Since the beginning of this article, the term "diffusion" is used regularly, because it is the most natural thing that can happen to monomers and dimers. For monomers it is obvious: they have no other possibility than to diffuse if they want to disappear (except to dissolve them in the substrate, which we will not consider here: we will work with a constant quantity of matter, or in other words, with constant number of atoms). The dimers can diffuse on the surface in the manner of monomers before going to meet another cluster already present. However, even if it seems natural to think that this mechanism describes what happens on the surface, one should not neglect other possibilities carrying out a theoretical study: the dimers could for example "decompose" leading to the diffusion of the monomers resulting from this "decomposition". There is also the last possibility (which I think unlikely): the dimer can "decompose", and the interaction of one of the monomers with the surface can be such that only one of the released atoms starts to diffuse.

Hereafter, I will call "diffusion" the process by which a dimer moves as a whole on the surface. The "dislocation" followed by the diffusion of two individual atoms will be called "explosion" and finally, the unlikely case where one of the atoms resulting from the "decomposition" remains fixed and that the other diffuse will be called "evaporation".

We note here that, in the case where the dimer diffuses, nothing prohibits that the monomers are stable on the surface. This case may seem really surprising, but in the calculations, this possibility can be considered without adding additional complications.

\section{Study assumptions.}
\label{hypotheses}
Six assumptions will be made for this study:
\begin{enumerate}
  \item Initial probabilities are known.
  \item Only dimers disappear.
  \item The diffusion of dimers or constituents can be decomposed in two of tree steps according the scenario:
  \begin{enumerate}
    \item If the dimer diffuses:
    \begin{enumerate}
        \item A dimer is removed from the set of dimers.
        \item The taken dimer is then placed randomly at the surface (i.e. on a nucleation site)
    \end{enumerate}
    \item If the dimer explode:
    \begin{enumerate}
        \item A dimer is removed from the set of dimers
        \item The first monomer is then placed randomly at the surface on a site holding a cluster.
        \item The second single atom is then also placed randomly on a site holding a cluster.
    \end{enumerate}
    \item If the dimer evaporates:
    \begin{enumerate}
        \item A dimer is removed from the set of dimers and a monomer is added to the set of monomers.
        \item A single atom is placed on a site on the surface.
    \end{enumerate}
  \end{enumerate}
  \item The number N of nucleation sites is very large : $1/N \ll 1$
  \item The mean free path of a diffusing atom is large in comparison of the distance of nucleation centers.
  \item Finally, the size of a cluster is negligible in regard of the distance of nucleation center. This means
        that the capture probability for an atom does not depend of the size of the cluster already present on
        the nucleation center, and that condition 5 will be fulfilled.
\end{enumerate}
\section{Dimer Diffusion}
\label{diffusion}

We will suppose here that the dimers leave their position and that, without breaking up, they diffuse until they reach a nucleation site (already occupied or not).

Let us consider a starting situation in which we have the probability $P_0$ of having an empty site, the probability $P_1$ of having a site with a monomer, $P_2$ of having a site with a dimer... and $P_n$ of having a site with a cluster of $n$ atoms. A priori, $P_1$ should be equal to 0, but, as said above, we are going to keep it, it won't make the calculations more complicated, and it could allow us to "solve" the improbable but not impossible problem where monomers, because of their enhanced interaction with the substrate, could be more efficiently trapped on the surface defects than the dimers.

The diffusion of the dimer can be split into two steps: 1 - a dimer is removed from the surface; 2 - the dimer is deposited on a randomly selected site. It will be noticed that the first half step of the diffusion will impact only the number of empty sites and the number of sites containing a dimer. The second step will affect the number of all other size classes.

Starting from \nameref{hypotheses}, the calculations given in \nameref{annexediffusion} lead to the following formulas summarizing the size statistics after dimer diffusion:

\color{blue}
\begin{equation}
  \begin{aligned}
   \underline{P_0} &= 1 - (1 - P_0 )e^{-x_a}  \\ 
    n~odd~:~
    \underline{P_{n \ne 0}} &= e^{-x_a} \left [ \frac{1}{(n/2)!}x_a^{n/2} (P_0 - 1) +
      \sum_{k=0}^{n/2-1}\frac{1}{k!}x_a^k{P_{n-2k}} \right ]\\ 
    n~even~:~
    \underline{P_n} &= e^{-x_a} \left [ \sum_{k=0}^{(n-1)/2}\frac{1}{k!}x_a^k{P_{n-2k}} \right ]\\
    with \\
    x_a &= \frac{P_2}{1 - P_0} 
  \end{aligned}
\label{diffusion_dimere} 
\end{equation}

\color{black}

$P_n$ being the initial probabilities before diffusion, and $\underline{P}_n$ the probabilities once all dimers have diffused and coalesced.

\section{Dimer explosion}
\label{explosion}

While less plausible than the case treated in section \ref{diffusion}, this case has to be solved, at least for the completeness of the present study. The differences between the size histograms calculated here and those calculated previously, when compared with experiments carefully carried out for this purpose, could allow to privilege this explosion mechanism over the diffusion mechanism, and provide indications on the hierarchy of energies involved in this or that system.

We suppose here that a dimer breaks up because of the temperature (or the chemical environment), and that the two monomers formed will diffuse very quickly to stick on a pre-existing nucleus.

It is important to realize that we cannot have stable monomers, and that if the two released monomers meet, it will be as if nothing had happened. We will therefore not consider this possibility.

Here the diffusion will take place in three steps: 1 - one dimer is removed from the surface; 2 - the first atom is dropped on a site containing a sufficiently stable cluster (at least a dimer); 3 - the second atom is also dropped on a sufficiently stable cluster.

The calculations detailed in \nameref{annexeexplosion} lead to the following formulas:

\color{blue}
\begin{equation}
\begin{split}
    \underline{P}_0 &= P_0 + x_a \\
    \underline{P}_1 &= 0 \\
    \underline{P}_2 &= 0 \\
    \underline{P}_{n\ge3} &= \left[ K_n + \sum_{i=1}^{n-2} K_{n-i} \frac{(-2)^i}{i!}[~ln(A-x_a)~]^i
    \right] (A-x_a)^2 - 2^{(n-2)} (A-x_a) \\
y    ~\\
    with~& \\
    x_a &= A - \frac{A^2}{A + P_2} \\
    K_{(n\ge3)} &= \frac{P_n+2^{n-2}A}{A^2} - \sum_{i=1}^{n-2} K_{n-i} \frac{(-2)^i}{i!}[~ln(A)~]^i  \\
    A &= 1-P_0\\
    and ~ K_2 &= \frac{A+P_2}{A^2}
\end{split}
\label{explosion_dimere}
\end{equation}
\color{black}

With always $P_n$ representing initial probabilities before the explosion, and $\underline{P}_n$ the probabilities once all dimers have vanished.

\section{Dimer evaporation}
\label{evaporation}

One can consider the situation in which one atom of the cluster is more strongly bound to the surface than to the other atom of the dimer. In this case, the thermal energy necessary to break a dimer can be too weak to activate the diffusion of one of the two atoms of the dimer: One of the atoms starts its random walk and the other remains fixed on the original site.

In this case, the diffusion splits in two steps: a - A dimer disappears and a monomer appears, b - The atom diffusing on the surface stops on a random site, be it empty, or occupied by a cluster of any size.

From the calculations described in \nameref{annexeevaporation}, we obtain the following formulas reflecting the evaporation of dimers.

\color{blue}
\begin{equation}
  \begin{aligned}
  \underline{P}_0 &= P_0 e^{-x}\\~\\
  \underline{P}_1 &= 1 - [1 - P_1 - x_aP_0] e^{-x_a} \\~\\
  \underline{P}_{n\ge2} &= \left[ \sum_{i=0}^n P_{n-i}\frac{x_a^i}{i!} {~}{~}
    - {~}{~} \frac{x_a^{(n-1)}}{(n-1)!} \right]  e^{-x_a}\\~\\
  ~\\
  with&\\ ~\\
  x_a &=  \frac{(1-P_1) - \sqrt{ (1-P_1)^2 - 2P_0P_2} }{P_0} ~ if ~ P_0 \ne 0 \\
  x_a &= \frac{P_2}{1-P_1} ~ if ~ P_0 = 0
  \end{aligned}
\label{evaporation_dimere}
\end{equation}
\color{black}

Again $P_n$ representing initial probabilities before the evaporation, and $\underline{P}_n$ the probabilities once all dimers have vanished.

\section{Discussion}

It is almost obvious that the dimers will "disappear" after the monomers have diffused and, unless a physical or chemical process is available to obtain a known size distribution on a surface with $P_1 = 0$, it will be necessary to use the formulas described in the paper on the diffusion of monomers\cite{sitja2021statistics} to handle the initial distribution before using the formulas given in this work.

\subsection{Experimental verification}

One of the first papers reporting the use of nano-structured surfaces for the organization of metal clusters gives enough details to make a comparison between an experimental result and the theoretical formulas obtained here. In this article published in 2006, the team of T. Michely uses a moiré produced by graphene on a surface (111) of an iridium (Ir) single crystal to organize Ir clusters. To know the deposited quantity they examine the 2D islands formed where no graphene is present, which allows to calibrate the deposit in a rather reliable way.\\
The table \ref{table1} allows to compare in a synthetic way the experimental results with the different theoretical predictions. As expected, the trivial formula \ref{formule_bidon} disagrees more as the deposit is low and as there are essentially monomers and dimers just after the deposit. The columns of the table labelled ``dimer diffusion'' and ``dimer explosion'' are calculated from the probabilities given by the formulas obtained in the reference \cite{sitja2021statistics}, which are themselves calculated from the Poisson distribution corresponding to the values $n$ of the average number of atoms per site. Here we see that the diffusion of dimers gives a result quite close to the explosion of dimers. We know anyway that in this case, it is the dimer diffusion that must be considered given the Ir-Ir and Ir-graphene interaction energies.\\
For each experimental point, we calculate the quantity of Ir that should have been deposited to obtain, with the theoretical formulas, the experimental coverage. This value is called "recalculated deposit". Here, it is the comparison of the recalculated deposit with the experimental deposit that is used to compare the experiment with the theory.
\begin{table}[h]
\begin{center}
\begin{tabular}{|c|c|c|c|c|c|c|}
  \hline
  Deposit   & $n$ & experimental    & formula \ref{formule_bidon} & dimer & recalculated/ &  dimer \\
 $\theta$ M.L. & ~   & $R_{occ}$  & ~   & diffusion \ref{diffusion_dimere}  & measured   & explosion \ref{explosion_dimere}  \\
  \hline
  0.01 & 0.87 & 19.2\%  &  5.8\%  & 19.4\% &  0.988 & 21.1\%  \\
  0.02 & 1.74 & 42.6\%  & 25.3\%  & 38.5\% &  1.109 & 40.9\%  \\
  0.03 & 2.61 & 65.5\%  & 48.4\%  & 56.2\% &  1.195 & 58.3\%  \\
  0.04 & 3.48 & 68.7\%  & 67.5\%  & 71.1\% &  0.954 & 72.3\%  \\
  0.05 & 4.35 & 82.9\%  & 80.9\%  & 82.2\% &  1.014 & 82.8\%  \\
  \hline
\end{tabular}
\captionsetup{width=.80\textwidth}
\caption{Comparison of experimental occupancy rates $R_{occ}$ with theoretical predictions for the different formulas. The formula \ref{formule_bidon} is the one reported in \cite{ISI:000366958900001}. The initial distribution before any diffusion is assumed to be given by the Poisson law. To easily compare the theoretical and experimental results, the $6^{th}$ column gives the ratio between theoretical deposit and the experimental deposit.}

\label{table1}
\end{center}
\end{table}

The comparison of the "calculated/measured" ratio in column 6 of table \ref{table1} shows that over the five experimental values, a very good agreement for three of them is obtained. The 0.02 monolayer deposit is in disagreement of 10\% and this disagreement rises to 20\% for the 0.03M.C. deposit. Can we incriminate measurement imprecision during the deposition for two of these experiments ? In fact it is difficult to say, however the experimenters are often confronted with experimental hazards not always controllable. The small difference between the occupancy rate between the deposit for 0.03M.C. and the following one for 0.04M.C. would tend to show that there is indeed an experimental under evaluation of the quantity deposited during the third deposit. To this we must add that in 2006 the aim of this work was not to confirm any theory but to prove that the type of substrate used made it possible to obtain particle arrays with a known distribution (the Poisson distribution) allowing to characterize properly the sample.
We can add that a wrong model would certainly give increasing (or decreasing) deviations with respect to the coverage, which is not the case here. The value for the deposit of 0.03ML distinguishes itself, but the other values are within the norm for a study that was not conducted with the aim of a rigorous comparison with a theory. One can hope that some experimenter will decide to accurately verify the conclusions of the formulas given here.

\subsection{Diffusion Vs Explosion}

While improbable (at least with atoms), one can envisage that experimentally a cluster would break before diffusing. One may wonder if the formulas allow distinguishing between one or the other of these scenarios. Since it is the dimers that will have a different behavior between both scenarios, one needs a distribution that before diffusion or explosion maximizes the number of dimers.
\begin{table}[h]
\begin{center}
\begin{tabular}{|c|c|c|c|}
  \hline
   ~ & Initial size distribution & dimer diffusion & dimer explosion \\
  \hline
  $P_0$               & 0.0      & 0.0 & 0.0  \\
  $P_1$               & 0.0      & 0.0 & 0.0  \\
  $P_2$               & $\alpha$ & 0.0 & 0.0  \\
  $P_3$               & 0.0      & 0.0 &  $0.1931\alpha$\\
  $P_4$               & 0.0      & $0.1839\alpha$ & $0.1736\alpha$ \\
  $P_5$               & 0.0      & 0.0 & $0.08876\alpha$ \\
  $P_6$               & 0.0      & $0.1226\alpha$ & $0.0324\alpha$ \\
  $P_7$               & 0.0      & 0.0 & $0.0093\alpha$ \\
  $P_8$               & 0.0      & $0.0459\alpha$ & $0.0022\alpha$ \\
  \hline
  Relative occupancy & 1 & $e^{-1} \simeq$ 0.36787 & 1/2 \\
  \hline
\end{tabular}
\captionsetup{width=.70\textwidth}
\caption{Comparative evolution of a dimer assembly under the effect of diffusion or explosion}
\label{table2}
\end{center}
\end{table}

If there are only dimers, the difference between diffusion and explosion is very important as shown in table \ref{table2}. If we can only access the occupancy rate, the ratio between the number of dimers deposited and the number of particles obtained gives a clear signal to discriminate one or the other mechanism: indeed, counting 37 clusters or 50 clusters in an area where there were 100 dimers should be easy even taking into account the uncertainty with only a single counting.
If experimentally one can count the number of atoms in the clusters obtained, the lack or the existence of clusters with an odd number of atoms is an infallible argument.

However, producing a sample with only dimers seems difficult unless one is depositing mass-sorted clusters. In most cases, the deposition of atoms is made by evaporation and condensation. One would suppose that in the case of our concern, one should make a deposit maximizing the number of dimers, but this is not the case, because the monomers present on the surface will certainly diffuse before the dimers. The results obtained by applying the formulas for monomer diffusion show that the ratio of dimers over the total number of clusters is maximized when the amount deposited tends towards 0. The size distribution after monomer diffusion\cite{sitja2021statistics} when $\theta \to 0$ is given by the following formula:

\begin{equation}
  \label{loidejojo}
  P_n = \frac{n-1}{n!}
\end{equation}

Taking as a starting point this distribution we obtain the occupancies listed in the table \ref{table3}. We notice that the differences are less important but not negligible: Counting on a zone where 1000 particles were initially found, we will find in one case $606 \pm 25$ clusters and in the other $666 \pm 25$ clusters.

\begin{table}[h]
\begin{center}
\begin{tabular}{|c|c|c|c|}
  \hline
   ~ &initial size distribution& dimer diffusion& dimer explosion\\
  \hline
  $P_0$               & 0       & 0 & 0  \\
  $P_1$               & 0       & 0 & 0  \\
  $P_2$               & 0.5       & 0.0 & 0.0  \\
  $P_3$               & 0.333333  & 0.2021768 & 0.2443238 \\
  $P_4$               & 0.125     & 0.1516326 & 0.2268339 \\
  $P_5$               & 0.033333  & 0.1213061 & 0.1239769 \\
  $P_6$               & 0.006944  & 0.0673922 & 0.0498425 \\
  $P_7$               & 0.0011904 & 0.0361029 & 0.0160665 \\
  $P_8$               & 0.0000144 & 0.0164087 & 0.0043355 \\
  \hline
  relative occupancy & 1 & 0.60653 & 0.66666 \\
  \hline
\end{tabular}
\captionsetup{width=.70\textwidth}
\caption{Comparative changes under diffusion or explosion from the size distribution given by formula \ref{loidejojo}.}
\label{table3}
\end{center}
\end{table}

\section{Conclusion}

While these formulae may be useful to discriminate different dimer disappearance scenarios, this is not their main use. If one considers a set of clusters giving properties to a given sample, optical, magnetic, or catalytic properties, the proper characterization of the proportion of clusters of such or such size is the only way to deduce the individual properties depending on the cluster size. By recording the response of a sample as a function of coverage, if monomer and dimer diffusion is involved, the formulas given in this study added to the formulas given in \cite{sitja2021statistics} allow measuring the activities of clusters of size equal to or larger than 3 as a function of their size. Although these formulas succeed in an experimental test, one can hope that they will be confirmed (or invalidated) by dedicated studies, which would allow them to be used without fear as is often done with the Poisson distribution.
Lastly, the main object of this study was related to the diffusion of atoms and dimers on a surface in the way it can occur in the systems studied in the field of nanosciences. Nevertheless, these formulas can equally be applied to other domains where atoms or molecules would not be on a surface but in a 3-dimensional matrix. The only importance is to respect the initial hypotheses (\nameref{hypotheses}).

\newpage

\bibliographystyle{../../../bibliographie/base_donnees/lefiltre.bst}
\bibliography{../../../bibliographie/base_donnees/biblio_jojo.bib}

\newpage
\section{Appendix 1 - Calculations for dimer diffusion}
\label{annexediffusion}

\subsection{Determining the recurrence formulas}
 
\subsubsection{$P_0$}
Let us consider a surface with a very large number N of nucleation centers. The number of empty sites $N_0$ is such that $N_0 = N \times P_0$. Removing a dimer is equivalent to increase $P_0$ by $1/N$. Indeed the number of empty sites after the first step is ${N_0}' = N_0 + 1$, $N_0 + 1 = {P_0}' \times N$ which means that ${P_0}' = {P_0} + 1/N$. In the rest of the document I will replace $1/N$ by $\varepsilon$ to simplify the notation.
The second step of the diffusion will consist of dropping randomly this dimer on one of the sites of the surface.
Obviously the probability that the dimer ``lands'' on a site containing n atoms is $P_n$. Here we are interested in knowing what happens to $P_0$. At the second stage of the diffusion $P_0$ is decreased by ${P_0}' \times \varepsilon$, i.e.: ${P_0}' = {}^{1}{P_0} = {P_0}' - {P_0}' \times \varepsilon$ and finally :

\[{}^{1}{P_0} = (P_0 + \varepsilon)(1-\varepsilon)\]

Noting ${}^{a}P_n$ the value of $P_n$ after $a$ elementary diffusions.

Trivially, we can extend this result. After a+1 diffusions :

\begin{equation}
{}^{a+1}{P_0} = ({}^{a}P_0 + \varepsilon)(1-\varepsilon)
\label{evolution_P0} 
\end{equation}

\subsubsection{$P_1$}

As mentioned earlier, we expect $P_1$ to be equal to 0 and remain so, but let's make the calculation as if the monomers were stable. The first step of the diffusion does not affect $P_1$, but the second step decreases the initial value of $P_1$ by $\varepsilon P_1$. In an immediate and trivial way, we can write that at the end of the first diffusion ${}^{1}{P_1} = P_1(1-\varepsilon)$. This leads to the following recurrence formula:

\begin{equation}
{}^{a+1}{P_1} = {}^{a}P_1(1-\varepsilon)
\label{evolution_P1} 
\end{equation}

\subsubsection{$P_2$}

The case of $P_2$ is a bit more tricky because the probability of having dimers at step $a+1$ depends on the probabilities $P_0$ and $P_2$ at step $a$. Obviously, the first step of the diffusion decreases the probability $P_2$ of $\varepsilon$ and we have ${P_2}' = {P_2}-\varepsilon$. The second step will increase it by $\varepsilon \times {P_0}'$ in the case that the dimer reaches an empty site, or decrease it by $\varepsilon \times {P_2}'$ if the dimer finishes its walk on a site occupied by another dimer. At the end of the second step of the diffusion, we have:

\[{}^{1}{P_2} = P_2(1-\varepsilon) - \varepsilon + \varepsilon P_0 + 2 \varepsilon^2\]

We can safely neglect the quadratic term in $\varepsilon$ and write the following recurrence formula:

\begin{equation}
{}^{a+1}{P_2} = {}^{a}P_2(1-\varepsilon) - \varepsilon(1 - {}^{a}P_0)
\label{evolution_P2} 
\end{equation}

\subsubsection{$P_3$}

$P_3$ is handled almost like $P_1$. Indeed the first step of the diffusion does not disturb the probability of having a trimer, but the second step will subtract 1 from $N_3$ in the situation where the dimer arrives on a site containing a trimer. There is another way to modify $P_3$, indeed if the dimer reaches a site containing a monomer we will add 1 to $N_3$. The new probability is consequently :

\[{}^1{P_3} = P_3(1-\varepsilon) + P_1\varepsilon\]

A formula that we will generalize to the following recurrence form:

\begin{equation}
{}^{a+1}{P_3} = {}^{a}P_3(1-\varepsilon) + \varepsilon{}^{a}P_1
\label{evolution_P3} 
\end{equation}

\subsubsection{$P_4$}

The only difference between $P_3$ and $P_4$ is that if $P_1$ was not altered by the first step of the diffusion, $P_2$ is. $P_4$ increases by $\varepsilon$ if the diffusing dimer meets another dimer.
The probability of this event is the number of dimers remaining after the first step of the diffusion, that is $P_2 - \varepsilon$. Finally, we can write:

\[{}^1{P_4} = P_4(1-\varepsilon) + (P_2 - \varepsilon)\varepsilon \]

i.e.:

\[{}^1{P_4} = P_4(1-\varepsilon) + \varepsilon P_2\]

once the quadratic terms removed.
A formula that will be extended to the following recurrence formula:

\begin{equation}
{}^{a+1}{P_4} = {}^{a}P_4(1-\varepsilon) + \varepsilon {}^{a}P_2
\label{evolution_P4} 
\end{equation}

\subsubsection{$P_5$ and $P_{n \geq 5}$ }

The evolution of $P_n$ for $n \geq 5$ and $P_3$ are strictly identical, in the sense that from size 5, the first step of the diffusion will have no impact on $P_5$: The evolution of $P_5$ depends on $P_5$ and $P_3$ and we can add that for a given size $n$ $P_n$ will only depend on $P_n$ and $P_{n-2}$. It follows:

\[{}^1{P_n} = P_n(1-\varepsilon) + \varepsilon P_{n-2}\]

That is extended to the following recurrence formula:

\begin{equation}
{}^{a+1}{P_n} = {}^{a}P_n(1-\varepsilon) + \varepsilon{}^{a}~P_{n-2}
\label{evolution_Pn} 
\end{equation}

\subsection{Iterations}

To derive the value of the different probabilities after a given number of single diffusions, one must iterate the recurrence formulas obtained previously.

\subsubsection{$P_0$}

Let's see how $P_0$ evolves. I recall the recurrence formula (\ref{evolution_P0}):

\[{}^{a+1}{P_0} = ({}^{a}P_0 + \varepsilon)(1-\varepsilon)\]

\[
\begin{aligned}
  {}^{a+1}{P_0} &= {}^{a}P_0(1-\varepsilon) + \varepsilon(1-\varepsilon) \\
               &= {}^{a}P_0(1-\varepsilon) + \varepsilon - \varepsilon^2
\end{aligned}  
\]

in which we can remove the quadratic terms in $\varepsilon$ :

\begin{equation}
{}^{a+1}{P_0} = {}^{a}P_0 + \varepsilon ( 1 -{}^{a}P_0 )
\label{evolution_P0_s} 
\end{equation}

The iterations result in:

\[
\begin{aligned}
  {}^{1}{P_0} &= P_0(1-\varepsilon) + \varepsilon \\
  {}^{2}{P_0} &= {}^{1}P_0(1-\varepsilon) + \varepsilon\\
  &= (P_0(1-\varepsilon) + \varepsilon)(1-\varepsilon) + \varepsilon\\
  &= P_0(1-\varepsilon)^2 + \varepsilon(1-\varepsilon) + \varepsilon\\
  {}^{3}{P_0} &= {}^{2}P_0(1-\varepsilon) + \varepsilon\\
  &= P_0(1-\varepsilon)^3 + \varepsilon(1-\varepsilon)^2 + \varepsilon(1-\varepsilon) + \varepsilon\\
  {}^{4}{P_0} &= {}^{3}P_0(1-\varepsilon) + \varepsilon\\
  &= P_0(1-\varepsilon)^4 + \varepsilon(1-\varepsilon)^3  + \varepsilon(1-\varepsilon)^2 + \varepsilon(1-\varepsilon) + \varepsilon
\end{aligned}  
\]

it seems to emerge a fairly simple pattern that would give :

\[
 {}^{a}{P_0} =  P_0(1-\varepsilon)^a + \varepsilon \sum _{i=0}^{a-1} (1-\varepsilon)^i 
 \]
 
We can easily prove that:
 
 \[
\sum _{i=0}^{a} \alpha^i = \frac { \alpha^{a+1} - 1 }{ \alpha - 1 }
\]

hence:

\[
\sum _{i=0}^{a-1} (1-\varepsilon)^i = \frac { (1-\varepsilon) ^a - 1 }{ (1-\varepsilon)  - 1 }
= \frac { 1 - (1-\varepsilon) ^a }{ \varepsilon}
\]

and finally:

\begin{equation}
{}^a{P_0} = 1 - (1 -P_0)(1-\varepsilon)^a 
\label{P0_discrete} 
\end{equation}

In the limit where $\varepsilon$ tends towards 0, we can deduce a continuous formula by noticing that $(1-\varepsilon)^a = e^{-a\varepsilon}$. The continuous version of the formula(\ref{P0_discrete}) is then:

\begin{equation}
P_0(x) = 1 - (1 - P_0 )e^{-x} 
\label{P0_continue} 
\end{equation}

writing $x=a\varepsilon$ which is nothing else than the number of diffusions per site ($a/N$)

\subsubsection{A short break to think}

We are interested here in the continuous versions of the evolution of probabilities. It is with them that we can easily do calculations. Obtaining $P_0(x)$ required quite a lot of work, and it would be good to ask ourselves if we could not get rid of the tedious discrete calculations to obtain an equivalent result, but in a simpler way.

For this, let us return to the formula (\ref{evolution_P0_s}) and see what it has to say. This formula is recalled here:

\[
  {}^{a+1}{P_0} = {}^{a}P_0 + \varepsilon ( 1 -{}^{a}P_0 )
\]

Each new diffusion increases $a$ by 1. Obviously, the number of elementary diffusions to reach a given state will be proportional to the number $N$ of sites considered, and we can introduce here the number $x$ of diffusions per site which is of course $a/N$. We have $x=a/N$, and making an additional diffusion increases $x$ by a small amount $\delta x$. Then we have: $x + \delta x = (a+1)/N$. It follows that $\delta x = \varepsilon$. We can add that $P_0(x)$ is also slightly modified by this additional diffusion. The formula (\ref{P0_simplifiee}) can thus be rewritten as follows:

\[
P_0(x) + \delta P_0 = P_0(x) +  ( 1 - P_0(x) ) \delta x
\]

Obviously in the limit where $\delta x$ tends to 0 we have :

\begin{equation}
  P_0(x) + \mathrm{d}P_0 = P_0(x) +  ( 1 - P_0(x) )\mathrm{d}x
\label{evol_P0_diff} 
\end{equation}

The equation (\ref{evol_P0_diff}) allows us to find a differential equation for which the solution should be what we are searching for:

\[
\frac{\mathrm{d}P_0}{\mathrm{d}x}(x) = 1 - P_0(x)
\]

i.e.: 
 
\begin{equation}
 P_0'(x) =   1 - P_0(x)  
\label{equadiff_P0} 
\end{equation}

Solving this equation is straightforward at this point, since we already know the solution that was computed earlier by the iteration method (\ref{P0_continue}). We are going nevertheless to check that this is the correct solution:

\[
\begin{aligned}
  P_0(x) &= 1 - (1 -P_0)e^{-x} \\
  P_0'(x) &= (1 - P_0) e^{-x}  \\~\\
  1 - P_0(x) &= 1 -  [1 - (1 -P_0)e^{-x}]\\
  &= (1 - P_0) e^{-x} = P_0'(x)
\end{aligned}
\]

which satisfies the differential equation (\ref{equadiff_P0}).

\subsubsection{$P_1$}

From the recurrence formula (\ref{evolution_P1}) for $P_1$ let's try to determine the differential equation that governs $P_1(x)$. We have:
\[
\begin{aligned}
{}^{a+1}{P_1} &= {}^{a}P_1 (1 - \varepsilon) \\
P_1(x) +  \mathrm{d}P_1(x) &=  P_1(x) - P_1(x)\mathrm{d}x \\
\frac{ \mathrm{d}P_1(x)}{\mathrm{d}x} &=  - P_1(x)
\end{aligned}
\]

That can simply be written:

\begin{equation}
 P_1'(x) = - P_1(x)  
\label{equadiff_P1} 
\end{equation}

The result is immediate:

\[
 P_1(x) = K_1 e^{-x}
 \]
 
for $x=0$ ; $P_1(0) = K_1 = P_1$ ; and finally:

\begin{equation}
P_1(x) = P_1 e^{-x} 
\label{P1_continue} 
\end{equation}

\subsubsection{$P_2$}

$P_2$ is a bit more complicated because it depends not only on $P_2$ but also on $P_0$. The recurrence formula (\ref{evolution_P2}) is arranged as follows:

\[
\begin{aligned}
{}^{a+1}{P_2} &= {}^{a}P_2 - {}^{a}P_2\varepsilon + {}^{a}P_0\varepsilon - \varepsilon \\
&=  {}^{a}P_2 + ( {}^{a}P_0 - {}^{a}P_2 - 1)\varepsilon
\end{aligned}
\]

The resulting differential equation is :

\[
 P_2'(x) = - P_2(x) + P_0(x) - 1
 \]
 
And since we know the solution for $P_0(x)$ (\ref{P0_continue}), we finally obtain the differential equation that we will attempt to solve:
 
\begin{equation}
 P_2'(x) = - P_2(x) - (1 - P_0 )e^{-x} 
\label{equadiff_P2} 
\end{equation}

We will assume that $P_2(x) = f(x)e^{-x}$. We then have:

\[
\begin{aligned}
  P_2'(x) &= f'(x)e^{-x} - f(x)e^{-x} \\
  &= (f'(x) - f(x))e^{-x}
\end{aligned}
\]

but (\ref{equadiff_P2}) leads to:

\[
\begin{aligned}
  P_2'(x) &= - P_2(x) - (1-P_0)e^{-x}\\
  &= - f(x)e^{-x} - (1-P_0)e^{-x} \\
  &= [ -f(x) - (1-P_0)]e^{-x}
\end{aligned}
\]

If the solution has the form that we assume, we must have:

\[
-f(x) - (1-P_0) = f'(x) - f(x)
\]
that is
\[
 f'(x) = - (1-P_0)
\]

it follows trivially :

\[
f(x) = -(1-P_0)x + K_2
\]

i.e.

\[
P_2(x) = [K_2 - (1-P_0)x ]e^{-x}
\]

As $P_2(0) = P_2$, we finally find:

\begin{equation}
P_2(x) =  [ P_2 - (1-P_0)x ]e^{-x}
\label{P2_continue} 
\end{equation}

\subsubsection{$P_3$}

$P_3$ is not more complicated than $P_2$. Indeed, from the recurrence formula (\ref{evolution_P3}) we derive the following differential equation:

\begin{equation}
 P_3'(x) = - P_3(x) + P_1(x)
\label{equadiff_P3} 
\end{equation}

We will once again assume that the solution has the following form $f(x)e^{-x}$, and it comes :

\[
P_3'(x) = (f'(x) - f(x))e^{-x} = (-f(x) + P_1)e^{-x}
\]

with the differential equation to solve:

\[
f'(x) = P_1
\]

the solution, once the limit for $x=0$ is included, is $f(x)=P_3 + xP_1$, from which we conclude:

\begin{equation}
P_3(x) =  [ P_3 + P_1x ]e^{-x}
\label{P3_continue} 
\end{equation}

\subsubsection{$P_4$}

The differential equation deduced from the formula (\ref{evolution_P4}) giving the evolution of $P_4$ is :

\[
P_4'(x) = - P_4(x) + P_2(x)
\]

Again using the same approach, $P_4(x) = f(x)e^{-x}$, we get :

\[
f'(x) = P_2 - ( 1 - P_0)x
\]

and finally:

\begin{equation}
P_4(x) =  [ P_4 +  P_2 x - \frac{1}{2}( 1 - P_0)x^2 ]e^{-x}
\label{P4_continue} 
\end{equation}

\subsubsection{$P_n$}

The point here is to understand why the generic solution $P_n(x) = f(x)e^{-x}$ works to find the solution to the differential equation from the formulas giving the evolutions of the probabilities.
the evolution of $P_n$ (\ref{evolution_Pn}) gives us the following differential equation :

\[
P_n'(x) = -P_n(x) + P_{n-2}(x) 
\]

Taking $P_n(x) = f_n(x)e^{-x}$, we will systematically obtain $P_n'(x) = (f_n'(x) - f_n(x))e^{-x}$, and if by chance, $P_{n-2}$ is such that $P_{n-2} = f_{n-2}(x)e^{-x}$, we will always have to solve:

\[
(f_n'(x) - f_n(x))e^{-x} = - f_n(x)e^{-x} + f_{n-2}(x)e^{-x}
\]

which will always be simplified to:

\begin{equation}
f_n'(x) = f_{n-2}(x)
\label{equadiff_generique} 
\end{equation}

which will have as a solution:

\begin{equation}
f_n(x) = P_n + \int_0^x {f_{n-2}(y)\mathrm{d}y }
\label{solution_generique} 
\end{equation}

In order to make $P_{n-2}(x)$ of the form $f_{n-2}(x)e^{-x}$, we need $P_{n-4}(x)$ to be of the form $f_{n-4}(x)e^{-x}$. And in order to make $P_{n-4}(x)$ of the form $f_{n-4}(x)e^{-x}$, we need $P_{n-6}(x)$ to be of the form $f_{n-6}(x)e^{-x}$.... this until we find a value of $n$ for which we know it is true. This is precisely the case since for $n=1$ and $n=2$ we have: $P_1(x) = P_1 e^{-x}$ and $P_2(x) =  [ P_2 - (1-P_0)x ]e^{-x}$.

From what has just been said, we can very simply give the successive $f_n(x)$ by integrating from one step to the next $f_{n-1}(x)$ :
\[
  \begin{aligned}
    f_2(x) &= P_2 - x(1-P_0) \\
    f_4(x) &= P_4 + xP_2 - \frac{x^2}{2}(1-P_0)\\
    f_6(x) &= P_6 + xP_4 + \frac{x^2}{2}P_2 - \frac{x^3}{6}(1-P_0)\\
    f_8(x) &= P_8    + xP_6 + \frac{x^2}{2}P_4 + \frac{x^3}{6}P_2 - \frac{x^4}{4!}(1-P_0)\\
    f_{10}(x) &= P_{10}+ xP_8 + \frac{x^2}{2}P_6 + \frac{x^3}{6}P_4 + \frac{x^4}{4!}P_2 - \frac{x^5}{5!}(1-P_0)\\
    f_{12}(x) &= P_{12}+ xP_{10}+\frac{x^2}{2}P_8 + \frac{x^3}{6}P_6 + \frac{x^4}{4!}P_4 + \frac{x^5}{5!}P_2 - \frac{x^6}{6!}(1-P_0)\\
    f_n(x) &= P_n + xP_{n-2} +  \frac{1}{2}x^2 P_{n-4} +  \frac{1}{6}x^3 P_{n-6} + \frac{1}{24}x^4 P_{n-8} + ...\\
                           &+ \frac{1}{(n/2 -1)!}x^{n/2 - 1} P_2 - \frac{1}{(n/2)!}x^{n/2} (1 - P_0)
   \end{aligned}
\]

and the probability for $n$ even can be summarized by:

\begin{equation}
  \begin{aligned}
    P_{n \ne 0}(x) &= e^{-x} \left [ \frac{1}{(n/2)!}x^{n/2} (P_0 - 1) + \sum_{k=0}^{n/2-1}\frac{1}{k!}x^k{P_{n-2k}} \right ]\\
    P_0(x) &= 1 - (1 - P_0 )e^{-x} 
  \end{aligned}
\label{solution_n_pair} 
\end{equation}

For $n$ odd, successive integrations of $f_n(x)$ lead to :

\[
  \begin{aligned}
    f_1(x) &= P_1 \\
    f_3(x) &= P_3 + x P_1 \\
    f_5(x) &= P_5 + x P_3 + \frac{1}{2} x^2 P_1 \\
    f_7(x) &= P_7 + x P_5 + \frac{1}{2} x^2 P_3 + \frac{1}{6} x^3 P_1 \\
    f_9(x) &= P_9 + x P_7 + \frac{1}{2} x^2 P_5 + \frac{1}{6} x^3 P_3 + \frac{1}{24} x^4 P_1 \\
    f_{11}(x) &= P_{11} + x P_9 + \frac{1}{2} x^2 P_7 + \frac{1}{6} x^3 P_5 + \frac{1}{4!} x^4 P_3 + \frac{1}{5!} x^5 P_1 \\
    f_n(x) &= P_n + xP_{n-2} +  \frac{1}{2}x^2 P_{n-4} +  \frac{1}{6}x^3 P_{n-6} + \frac{1}{24}x^4 P_{n-8} + ...\\
                           &+ \frac{1}{[(n-1)/2 -1]!}x^{(n-1)/2 - 1} P_3 + \frac{1}{[(n-1)/2]!}x^{(n-1)/2} P_1
   \end{aligned}
\]

and the probability for $n$ odd can be summarized by:

\begin{equation}
    P_n(x) = e^{-x} \left [ \sum_{k=0}^{(n-1)/2}\frac{1}{k!}x^k{P_{n-2k}} \right ]
\label{solution_n_impair} 
\end{equation}

\subsection{Diffusion termination}

Of course, as we are dealing here with the diffusion of dimers, the process will stop when the number of dimers is equal to 0. This is summarized by the condition $P_2(x) = 0$. The value of $x_a$ for which $P_2(x_a) = 0$ is easily calculated:

\begin{equation}
    x_a = \frac{P_2}{1 - P_0}
\label{arret_cas1} 
\end{equation}

\subsection{Diffusion summary.}

To conclude, the size distribution after dimer diffusion is summarized in the following formulas:

\color{blue}
\begin{equation}
  \begin{aligned}
   \underline{P_0} &= 1 - (1 - P_0 )e^{-x_a}  \\ ~ \\
    n~pair~:~
    \underline{P_{n \ne 0}} &= e^{-x_a} \left [ \frac{1}{(n/2)!}x_a^{n/2} (P_0 - 1) +
      \sum_{k=0}^{n/2-1}\frac{1}{k!}x_a^k{P_{n-2k}} \right ]\\ ~ \\
    n~impair~:~
    \underline{P_n} &= e^{-x_a} \left [ \sum_{k=0}^{(n-1)/2}\frac{1}{k!}x_a^k{P_{n-2k}} \right ] \\ ~ \\
    with&  \\ ~ \\
    x_a &= \frac{P_2}{1 - P_0}
  \end{aligned}
\label{diffusion_en_bloc} 
\end{equation}

\color{black}

\newpage
\section{Appendix 2 - Calculations for dimer explosion}
\label{annexeexplosion}

\subsection{Determination of increments, of differential equations, and of probabilities}

\subsubsection{$P_0$}

The first step of the diffusion increases $P_0$ by $\varepsilon$. The following steps do not influence $P_0$, since once deposited, the atom will diffuse very quickly to an already occupied site. So we get :

\[
\begin{aligned}
  P_0' &= P_0 + \varepsilon \\
  P_0'' &= P_0' \\
  P_0''' &= P_0''
\end{aligned}
\]

that is:

\[
{}^1P_0 =  P_0 + \varepsilon
\]

The differential equation is:

\begin{equation}
    P_0'(x) = 1
\label{equadiff_eP0} 
\end{equation}

$P_0(x)$ is very easy to find in this case:

\begin{equation}
    P_0(x) = P_0 + x
\label{eP0} 
\end{equation}

\subsubsection{$P_1$}

Here, monomers are definitely prohibited, and no matter what happens :

\begin{equation}
    P_1(x) = 0
\label{eP1} 
\end{equation}

\subsubsection{$P_2$}

$P_2$ is affected by each of the three steps of the diffusion: Step 1 removes $\varepsilon$ from $P_2$ the two following steps add $\varepsilon$ according to the probability of the released monomers to fall on a dimer. One must be extremely careful here because the probability that an atom lands on a dimer is not equal to the probability of having a dimer ($P_2$): Indeed, in the case we are interested in here, putting an atom on an empty site does not make sense, it can only stabilize while waiting for the next diffusion on a site already occupied. After the first step of the diffusion, the number of occupied sites is $(1 - P_0') = 1 - P_0 - \varepsilon$. The sequence of the three diffusion steps thus modifies the probability of having a dimer as follows:

\[
\begin{aligned}
  P_2'   &= P_2 - \varepsilon \\
  P_2''  &= P_2' - \frac{P_2'\varepsilon}{1 - P_0 - \varepsilon } \\
  P_2''' &= P_2'' - \frac{P_2''\varepsilon}{1 - P_0 - \varepsilon }
\end{aligned}
\]

Once expanded and negligible terms dropped we obtain :

\[
  {}^1P_2 =  P_2 - \frac{2 P_2}{1-P_0} \varepsilon - \varepsilon
\]
  
Resulting in the following differential equation:

\[
  P_2'(x) = -\frac{2P_2(x)}{1 - P_0(x)} - 1
\]
  
which becomes when replacing $P_0(x)$ given by the formula (\ref{eP0}) :

\begin{equation}
  P_2'(x) = -\frac{2P_2(x)}{1 - P_0 - x} - 1
\label{equadiff_eP2} 
\end{equation}

To simplify the notation, we will assume here and for the following $1-P_0=A$. We can rewrite the previous equation:

\begin{equation}
  P_2'(x) = -\frac{2P_2(x)}{A-x} - 1
\label{equadiff_eP2A} 
\end{equation}

This differential equation is not easy to solve. One can for example iterate the individual diffusions to give clues. An other method, more direct, is to note that since we have the term $-\frac{2P_2(x)}{A - x}$ in the equation, a solution that could work would be of the form:

\[
P_2(x)=f_2(x)(A-x)^2
\]

so:

\[
\begin{aligned}
  P_2'(x) &= (A-x)^2f_2'(x)- 2(A-x)f_2(x)\\
  &=(A-x)^2f_2'(x) - \frac{2P_2(x)}{A-x}
\end{aligned}
\]

The equation (\ref{equadiff_eP2A}) results in :

\[
(A-x)^2f_2'(x)=-1
\]

that is to say :

\[
f_2'(x)=-(A-x)^{-2}
\]

leading to 

\[
f_2(x)=-(A-x)^{-1} + K_2
\]

hence

\[
P_2(x)=-(A-x) + K_2(A-x)^2
\]

We must have $P_2(0)=P_2$, which determines the constant $K_2$ :

\[
  K_2A^2-A=P_2
\]
\[
  \Longrightarrow
\]
\begin{equation}
  K_2=\frac{A+P_2}{A^2}
\label{K2} 
\end{equation}

finally

\begin{equation}
   P_2(x)=\frac{A+P_2}{A^2}(A-x)^2-(A-x)
\label{eP2A} 
\end{equation}

And substituting $A$ with $1-P_0$ :

\begin{equation}
  P_2(x) = \frac{1 - P_0 + P_2}{(1 - P_0)^2}( 1 - P_0 - x)^2 - (1 - P_0 - x )
\label{eP2} 
\end{equation}

The verification of this solution can be found in paragraph \nameref{peP2} in the supplementary information.

We can now know when the diffusion of dimers will end, obviously, there are two solutions to annul $P_2(x)$ :

\[
x_1=A
\]

and

\[
x_2 = A( 1 - \frac{A}{A+P_2})
\]

The first solution seems not very realistic since it depends only on $P_0$, and we expect it, at least, to depend on $P_2$.
Note that $A \ge 0$ and $P_2 \ge 0$ leading to $[1-A/(A+P2)] \le 1$ . $x_2$ is then necessarily smaller than $x_1$ since it is the result of $x_1$ multiplied by a factor smaller than 1. $x_2$ will be the first value to be reached during the explosion process, and we can forget about the solution $x_1$ which is obviously not a physical solution. $x_2$ has the expected behaviour: it starts from $0$ when $P_2 = 0$ and grows by increasing $P_2$. This is the solution we shall use to determine the end of the "diffusion" of the dimers.

\color{blue}
\begin{equation}
 x_a = (1 - P_0)( 1 - \frac{1 - P_0}{1 - P_0 + P_2}) 
\label{earret} 
\end{equation}
\color{black}

\subsubsection{$P_3$}

The 3 steps of diffusion alter the probability of having a trimer in the following way:

\[
\begin{aligned}
  P_3'   &= P_3 \\
  P_3''  &= P_3' - \frac{P_3'\varepsilon }{1 - P_0 - \varepsilon } + \frac{P_2'\varepsilon }{1 - P_0 - \varepsilon }\\
  P_3''' &= P_3'' - \frac{P_3''\varepsilon }{1 - P_0 - \varepsilon } + \frac{P_2''\varepsilon }{1 - P_0 - \varepsilon }
\end{aligned}
\]

Once all the negligible terms removed, we obtain the following differential equation:

\[
P_3(x)' = \frac{2}{1-P_0(x)} \left( P_2(x) - P_3(x) \right )
\]

substituting $P_0(x)$ and $P_2(x)$ with the previously determined expressions (\ref{eP0}) and (\ref{eP2}), and still substituting $1-P_0$ with $A$ we get :

\[
P_3'(x) = \frac{2}{A - x} \left [  \frac{A+P_2}{A^2}(A-x)^2-(A-x)  \right  ] - 2 \frac{P_3(x)}{A - x}
\]

or,once reduced:

\begin{equation}
  P_3'(x) = 2 \left [  \frac{A + P_2}{A^2}(A - x ) - 1 \right ] - 2\frac{P_3(x)}{A - x }
\label{equadiff_eP3} 
\end{equation}

We note that $P_3'(x)$ depends on $2P_3(x)/(A-x)$ which leads us, as for $P_2(x)$, to a solution of the form: $P_3(x) = f_3(x)(A-x)^2$.

\[
\begin{aligned}
  P_3'(x) &= (A-x)^2f_3'(x)- 2(A-x)f_3(x)\\
  &=(A-x)^2f_3'(x) - \frac{2P_3(x)}{A-x}
\end{aligned}
\]

thus:

\[
\begin{aligned}
  &(A-x)^2f_3'(x) =  2 \left [  \frac{A + P_2}{A^2}(A - x ) - 1 \right ]\\
  &f_3'(x) =  2 \frac{A + P_2}{A^2} \frac{1}{A - x} -  \frac{2}{(A - x)^2}\\
  &f_3(x) =  -2 \frac{A + P_2}{A^2} ln(A - x) -  \frac{2}{(A - x)} + K_3
\end{aligned}
\]

leading to:

\[
\begin{aligned}
  P_3(x) &=f_3(x)(A-x)^2 \\
    &=(A-x)^2 \left [  -2 \frac{A + P_2}{A^2} ln(A - x) -  \frac{2}{(A - x)} + K_3 \right ]\\
    &= -2 \frac{A + P_2}{A^2} ln(A - x)(A-x)^2  - 2(A - x) + K_3(A-x)^2 
\end{aligned}
\]

as $P_3(0) = P_3$

\[
  P_3 = -2 \frac{A + P_2}{A^2} ln(A)A^2  - 2A + K_3A^2 \\
  K_3 = \frac{ P_3 + 2A}{A^2} +  2\frac{(A + P_2)}{A^2}ln(A)
  \]
  
\begin{equation}
  K_3 = \frac{ P_3 + 2A}{A^2} +  2\frac{(A + P_2)}{A^2}ln(A)
\label{K3} 
\end{equation}

finally:

\[
\begin{aligned}
  P_3(x) &= -2 \frac{A + P_2}{A^2} ln(A - x)(A-x)^2  - 2(A - x) +  \frac{ P_3 + 2A}{A^2}(A-x)^2
  + 2\frac{(A + P_2)}{A^2}ln(A)(A-x)^2\\~\\
  &= -2 \frac{A + P_2}{A^2} ln(A - x)(A-x)^2 +\frac{ P_3 + 2A}{A^2}(A-x)^2
  + 2\frac{(A + P_2)}{A^2}ln(A)(A-x)^2  - 2(A - x) \\~\\
&=  \left [ -2 \frac{A + P_2}{A^2} ln(A - x) +\frac{ P_3 + 2A}{A^2}
  + 2\frac{(A + P_2)}{A^2}ln(A) \right](A-x)^2  - 2(A - x) \\~\\
  &=  \left [ \frac{ P_3 + 2A}{A^2}  + 2\frac{(A + P_2)}{A^2}ln(A)
    - 2 \frac{A + P_2}{A^2} ln(A - x) \right](A-x)^2 - 2(A - x) \\~\\
&=  \left [ \frac{ P_3 + 2A}{A^2} + 2 \frac{A + P_2}{A^2}({~}ln(A)- ln(A - x){~})  \right](A-x)^2
  - 2(A - x) \\~\\
 \end{aligned}
\]
\\~\\~
that is :
\\~\\~
\begin{equation}
\begin{split}
  P_3(x) &= \left [ \frac{ P_3 + 2A}{A^2} + 2 \frac{A + P_2}{A^2}({~}ln(A)- ln(A - x){~})  \right](A-x)^2
  - 2(A - x)
   \\~\\
  with ~A &= 1-P_0
\label{eP3} 
\end{split}
\end{equation}

One can be convinced of the validity of this formula looking at the section : \nameref{epP3} in the supplementary materials\\
~\\
Before going further we can rearrange this solution in terms of $ln(A-x)$; $(A-x)$; and $(A-x)^2$ :

\[
\begin{aligned}
  P_3(x) &= -2 \frac{A + P_2}{A^2} ln(A - x)(A-x)^2  - 2(A - x) +  \frac{ P_3 + 2A}{A^2}(A-x)^2
  + 2\frac{(A + P_2)}{A^2}ln(A)(A-x)^2\\~\\
 &=\left[  \frac{ P_3 + 2A}{A^2}
  + 2\frac{(A + P_2)}{A^2}ln(A)\right] (A-x)^2  - 2 \frac{A + P_2}{A^2} ln(A - x)(A-x)^2   - 2(A - x) \\~\\
 \end{aligned}
\]

Recalling the values of $K_2$ (\ref{K2}) and $K_3$ (\ref{K3}) we obtain:

\begin{equation}
 P_3(x)= K_3 (A-x)^2  - 2 K_2 ln(A - x)(A-x)^2   - 2(A - x) \\~\\
 \label{eP3AK} 
\end{equation}

\subsubsection{$P_4$ and $P_{n\ge3}$}

The 3 steps of the diffusion change the probability of having an cluster of 4 atoms in the following way:

\[
\begin{aligned}
  P_4'   &= P_4 \\
  P_4''  &= P_4' - \frac{P_4'\varepsilon }{1 - P_0'} + \frac{P_3'\varepsilon }{1 - P_0'}\\
  P_4''' &= P_4'' - \frac{P_4''\varepsilon }{1 - P_0'} + \frac{P_3''\varepsilon }{1 - P_0'}
\end{aligned}
\]

Once all the negligible terms are cleared, we obtain the following differential equation:

\[
P_4(x)' = \frac{2}{1-P_0(x)} \left[ P_3(x) - P_4(x) \right]
\]

We will notice first that, starting from $P_3$, the differential equations are all of the same form:

\begin{equation}
\begin{split}
P_n(x)' = \frac{2 P_{n-1}(x)}{A-x} - \frac{2 P_n(x)}{A-x}
\label{edifgenerique}
\end{split}
\end{equation}

We will also remark that assuming that $P_n(x)=f_n(x)(A-x)^2$ the term $-\frac{2 P_n(x)}{A-x}$ appears in $P_n'(x)$:

\begin{equation}
\begin{split}
  P_n(x)' &= \left[ f_n(x)(A-x)^2 \right]' =  f_n'(x)(A-x)^2 - 2 f_n(x)(A-x) \\
  &=  f_n'(x)(A-x)^2 -  \frac{2 f_n(x)(A-x)^2}{A-x} \\
  &=  f_n'(x)(A-x)^2 -  \frac{2 P_n(x)}{A-x}
\label{solutiongenerique}
\end{split}
\end{equation}

From (\ref{edifgenerique}) and (\ref{solutiongenerique}) it follows :

\[
f_n'(x)(A-x)^2 = \frac{2 P_{n-1}(x)}{A-x}
\]

that is :

\[
f_n'(x)(A-x)^2 = \frac{2 f_{n-1}(x)(A-x)^2}{A-x}
\]
\\~
\\~
which yields to the following relationship for $n\ge3$ :
\\~
\begin{equation}
f_n'(x) = \frac{2 f_{n-1}(x)}{A-x}
\label{edifgenerique+}
\end{equation}
\\~
\\~
Starting from $f_2(x)$, multiplying by  $\frac{2}{(A-x)}$ and integrating, will give us the solutions we are searching for, and this from the size 2 for which the generic differential equation (\ref{edifgenerique}) starts to apply. We will make it simpler by noting that :

\[
\left([ln(A-x)]^n\right)' = -n\frac{[ln(A-x)]^{n-1}}{A-x}
\]

and then:

\[
\int{\frac{[ln(A-x)]^{n}}{A-x}} = -\frac{1}{n+1}[ln(A-x)]^{n+1}
\]
\\~
\\~
and also, in a less stressful way:

\[
\int{\frac{1}{(A-x)^2}} = \frac{1}{A-x}
\]
\\~
\\~
Thus, the solution sequence begins as follows:\\~
\\~
\[
\begin{aligned}
  f_2(x) &= K_2-\frac{1}{A-x}\\
\\
  2\frac{f_2(x)}{A-x} &=  \frac{2K_2}{A-x} - \frac{2}{(A-x)^2}\\
  f_3(x) &= K_3 + \int{ \frac{2K_2}{A-x}} - \int{\frac{2}{(A-x)^2}}\\
 &=  K_3 - 2K_2ln(A-x) - \frac{2}{A-x}\\
\\
  2\frac{f_3(x)}{A-x} &=  \frac{2K_3}{A-x} - \frac{4K_2ln(A-x)}{A-x}- \frac{4}{(A-x)^2}\\
  f_4(x) &= K_4 + \int{ \frac{2K_3}{A-x}} - \int{ \frac{4K_2ln(A-x)}{A-x}} -  \int{\frac{4}{(A-x)^2}}\\
&= K_4  - 2K_3ln(A-x) +\frac{4K_2[ln(A-x)]^2}{2!} -  \frac{4}{A-x}\\
  \\
  2\frac{f_4(x)}{A-x} &= \frac{2K_4}{A-x} - \frac{4K_3ln(A-x)}{A-x}
  +\frac{8[K_2ln(A-x)]^2}{2!(A-x)} - \frac{8}{(A-x)^2} \\
  f_5(x) &= K_5 + \int{\frac{2K_4}{A-x}} -  \int{\frac{4K_3ln(A-x)}{A-x}}
  + \int{\frac{8K_2[ln(A-x)]^2}{2!(A-x)}} -  \int{\frac{8}{(A-x)^2}} \\
&= K_5 - 2K_4ln(A-x) + \frac{4K_3[ln(A-x)]^2}{2!} - \frac{8K_2[ln(A-x)]^3}{3!} -  \frac{8}{A-x}\\
\end{aligned}
\]
\\~
\\~
  
We notice that from one step to the other the term  $[ln(A-x)]^n]$ turns into $\frac{-2}{(n+1)}[ln(A-x)]^{n+1}$, and that the term  $\frac{1}{A-x}$ turns into $\frac{2}{A-x}$.
  This allows us to obtain the following general law:
  
\begin{equation}
  f_{(n\ge3)}(x) = K_n + \left[ \sum_{i=1}^{n-2} K_{n-i} \frac{(-2)^i}{i!}[~ln(A-x)~]^i \right] - \frac{2^{(n-2)}}{A-x}
\label{formulegenerale}
\end{equation}

Knowing that $P_n=P_n(0)=f_n(0)A^2$ we deduce the value of the constants :

\begin{equation}
  K_{(n\ge3)} = \frac{P_n+2^{n-2}A}{A^2} - \sum_{i=1}^{n-2} K_{n-i} \frac{(-2)^i}{i!}[~ln(A)~]^i 
\label{constantegenerale}
\end{equation}

The correctness of the general formula (\ref{formulegenerale}) is proved in the section \nameref{pePn} of supplementary materials.

\subsection{Explosion recap}

In the end, we will summarize in these few following lines the statistics after the "explosion" of all dimers:

\color{blue}
\begin{equation}
\begin{split}    ~\\
    \underline{P}_0 &= P_0 + x_a \\
    \underline{P}_1 &= 0 \\
    \underline{P}_2 &= 0 \\
    \underline{P}_{n\ge3} &= \left[ K_n + \sum_{i=1}^{n-2} K_{n-i} \frac{(-2)^i}{i!}[~ln(A-x_a)~]^i
    \right] (A-x_a)^2 - 2^{(n-2)} (A-x_a) \\~\\
    with~& \\~\\
    A &= 1-P_0~;\\
    K_{(n\ge3)} &= \frac{P_n+2^{n-2}A}{A^2} - \sum_{i=1}^{n-2} K_{n-i} \frac{(-2)^i}{i!}[~ln(A)~]^i~; \\
    K_2 &= \frac{A+P_2}{A^2}~; \\~\\
    and~&\\~\\
    x_a &= A - \frac{A^2}{A + P_2}
\end{split}
\label{explostat}
\end{equation}
\color{black}

Where the values of $P_n(x_a)$ are written $\underline{P}_n$.

\newpage
\section{Appendix 3 - Calculations for dimer evaporation}
\label{annexeevaporation}

\subsection{Modification of the probabilities}

In the same manner as we did in the two previous sections, we will determine the variations of the probabilities according to $\varepsilon$ and derive laws that apply to all sizes. Always noting <<~'~>> the situation after the first step and <<~''~>> the situation after the second step. The situation is the same as for ``diffusion'' in the sense that, as all sites are eligible to receive a monomer, there will not be the decrease in the possibilities that introduced the logarithms in the case of the "explosion" of dimers. \\The transformations of the probabilities can be summarized as follows:

\begin{equation}
\begin{split}
  P_0 ~ \longrightarrow ~ P_0' &= P_0 ~ \longrightarrow ~ P_0'' = P_0 - \varepsilon P_0\\
  P_1 ~ \longrightarrow ~ P_1' &= P_1 + \varepsilon  ~ \longrightarrow ~ P_1''
  = P_1 + \varepsilon - \varepsilon (P_1 +  \varepsilon) + \varepsilon P_0 \\
  P_2 ~ \longrightarrow ~ P_2' &= P_2 - \varepsilon  ~ \longrightarrow ~ P_2''
  = P_2 - \varepsilon - \varepsilon (P_2 -  \varepsilon) + \varepsilon (P_1 + \varepsilon) \\
  P_3 ~ \longrightarrow ~ P_3' &= P_3  ~ \longrightarrow ~ P_3''
  = P_3 - \varepsilon P_3+ \varepsilon (P_2 - \varepsilon) \\
  P_4 ~\longrightarrow~ P_4' &= P_4  ~\longrightarrow~ P_4'' = P_4 - \varepsilon P_4+ \varepsilon P_3 \\
  P_n ~\longrightarrow~ P_n' &= P_n  ~\longrightarrow~ P_n'' = P_n - \varepsilon P_n+ \varepsilon P_{n-1} \\
\end{split}
\label{etapevap1}
\end{equation}

Discarding the negligible terms we have:

\begin{equation}
\begin{split}
  P_0'' &= P_0 - \varepsilon P_0\\
  P_1'' &= P_1 + \varepsilon - \varepsilon P_1 + \varepsilon P_0 \\
  P_2'' &= P_2 - \varepsilon - \varepsilon P_2 + \varepsilon P_1 \\
  P_3'' &= P_3 - \varepsilon P_3 + \varepsilon P_2 \\
  P_n'' &= P_n - \varepsilon P_n + \varepsilon P_{n-1} \\
\end{split}
\label{etapevap2}
\end{equation}

\subsection{Differential equation and their solutions}

From the formulas (\ref{etapevap2}) one deduces the following differential equations:

\begin{equation}
\begin{split}
  P_0'(x) &= -P_0(x)\\
  P_1'(x) &= 1 + P_0(x) - P_1(x)\\
  P_2'(x) &= -1 + P_1(x) - P_2(x)\\
  P_3'(x) &= P_2(x) - P_3(x)\\
  P_n'(x) &= P_{n-1}(x) - P_n(x)\\
\end{split}
\label{etapevap2}
\end{equation}

Here we are exactly in the same case as in the section \nameref{diffusion}. We have the derivative $P_n'(x)$ which involves $-P_n(x)$ and the natural solution is $P_n(x) = f_n(x)e^{-x}$. In the same way as in the section \nameref{diffusion}, we will have:

\[
  P_n'(x) = f_n'(x)e^{-x} - P_n(x)
\]

These differential equations are easily solved:

\subsubsection{$P_0$}

\[
  P_0'(x) = f_0'(x)e^{-x} - P_0(x) =  -P_0(x) \Leftrightarrow  f_0'(x)e^{-x} = 0
\]

The obvious solution is $f_0(x) = P_0$ and then we have:

\begin{equation}
  P_0(x)=P_0 e^{-x}
\label{evaP0}
\end{equation}

\subsubsection{$P_1$}
We have to solve :
\[
  P_1'(x) = 1 + P_0(x) - P_1(x)
\]
assuming that  $P_1(x) = f_1(x)e^{-x}$, The differential equation For $P_1$ can be summarized as:

\[
 f_1'(x)e^{-x} =  1 + P_0(x)
\]

that is not difficult to solve:

\[
  \begin{aligned}
    f_1'(x)e^{-x} &=  1 + P_0 e^{-x}\\
    f_1'(x) &= e^x +  P_0
  \end{aligned}
\]

leading trivially to:

\[
f_1(x) = K_1 +  e^x +  xP_0
\]

and

\[
P_1(x) = K_1 e^{-x} +  1  +  xP_0e^{-x}
\]

Of course, $K_1$ is such as $P_1(0)=P_1$

\[
  \begin{aligned}
    P_1 &= K_1+  1 \\
    K_1 &= P_1 - 1
  \end{aligned}
\]

finally:

\begin{equation}
  P_1(x)= 1 - [1 - P_1 - xP_0] e^{-x} 
\label{evaP1}
\end{equation}

\subsubsection{$P_2$}
For $P_2$, we have:
\[
  \begin{aligned}
  P_2'(x) &= -1 + P_1(x) - P_2(x)\\~\\
    f_2'(x)e^{-x} &=  -1 + P_1(x)\\
    f_2'(x) &= -e^x + \left \{ 1 - [1 - P_1 - xP_0] e^{-x}  \right \} e^x\\
    &= -e^x + e^x  - [1 - P_1 - xP_0] \\
    &= P_1 + xP_0 - 1\\
  \end{aligned}
\]

In a trivial way :

\[
   f_2(x) = K_2 + xP_1 + \frac{1}{2}x^2 P_0 - x
\]

and simply $K_2 = P_2$, which results in:

\begin{equation}
  P_2(x)=\left[  P_2  + xP_1 + \frac{1}{2}x^2 P_0  - x \right] e^{-x} 
\label{evaP2}
\end{equation}

\subsubsection{$P_{n\ge3}$}

From $P_3$, the differential equations are all the same: ($P_n'(x) = P_{n-1}(x) - P_n(x)$).
Moreover, $P_2(x)$ has the form of a simple polynomial in powers of $x$ multiplied by a factor \( e^{-x} \). The solution: \( f_n'(x)e^{-x} = \mathcal{P}_{n-1}(x)e^{-x} \) will always lead to \( f_n'(x) = \mathcal{P}_{n-1}(x) \) which require $f_n(x)$ to be a polynomial as well: $f_n(x) = \mathcal{P}_n(x)$.
The successive solutions of $f_n(x)$ are therefore polynomials whose constant will necessarily be equal to $P_n$ since all the other terms are factors of a power of $x$ and that $e^0=1$.

\begin{equation}
  \begin{split}
    P_2(x) = \left[  P_2 {~}{~}{~} + {~}{~}{~} xP_1{~}{~} +{~}{~} \frac{1}{2}x^2 P_0 {~} - {~} x {~}{~}{~} \right] e^{-x}\\
    P_3(x) = \left[  P_3 {~}{~}{~} + {~}{~}{~} xP_2{~}{~} +{~} \frac{1}{2}x^2 P_1 {~} + {~}\frac{1}{3!}x^3 P_0 {~}
      -{~}  \frac{1}{2}x^2 \right] e^{-x}\\
    P_4(x) = \left[  P_4{~}  +{~} xP_3{~} +{~} \frac{1}{2}x^2 P_2{~}  + {~}\frac{1}{3!}x^3 P_1
     {~} +{~} \frac{1}{4!}x^4 P_0 {~} - {~} \frac{1}{3!}x^3 \right] e^{-x}
  \end{split}
\label{evaP3}
\end{equation}

And the appropriate generalization for $P_n(x)$ is:

\begin{equation}
    P_{n\ge2}(x) = \left[ \sum_{i=0}^n P_{n-i}\frac{x^i}{i!} {~}{~} - {~}{~} \frac{x^{(n-1)}}{(n-1)!} \right]  e^{-x}
\label{evaPn}
\end{equation}

\subsection{halting evaporation}

Dimer evaporation stops naturally when all dimers have finally disappeared. This condition is achieved when $P_2(x) = 0$.

\begin{equation}
  P_2(x)=0 \Leftrightarrow  P_2 + xP_1 + \frac{1}{2}x^2 P_0  -  x = 0
\label{evap2=0}
\end{equation}

There are two solutions to annul $P_2(x)$

\[
  x=\frac{(1-P_1) \pm \sqrt{ (1-P_1)^2 - 2P_0P_2} }{P_0}
\]

We will notice two things. First, $1-P_1 \ge 0$ and $P_0P_2 \ge 0$, and if $(1-P_1)^2 - 2P_0P_2 \ge 0$, then $\sqrt{ (1-P_1)^2 - 2P_0P_2}  \le (1-P_1) $. From these remarks it follows that:

\[
  \frac{(1-P_1) - \sqrt{ (1-P_1)^2 - 2P_0P_2} }{P_0} \le \frac{(1-P_1) + \sqrt{ (1-P_1)^2 - 2P_0P_2} }{P_0}
\]

The solution we are looking for is therefore the first one encountered, i.e.:

\begin{equation}
  x_a =  \frac{(1-P_1) - \sqrt{ (1-P_1)^2 - 2P_0P_2} }{P_0} 
\label{evaparret}
\end{equation}

Note that if $P_0$ tends towards 0, then we can either do a Taylor expansion of $\sqrt{ (1-P_1)^2 - 2P_0P_2}$:

\[
  \lim_{P_0 \to 0}  \sqrt{ (1-P_1)^2 - 2P_0P_2} = (1-P_1) - \frac{2P_0P_2}{2 \sqrt{(1-P_1)^2}}
  =(1-P_1) - \frac{P_0P_2}{(1-P_1)}
\]

leading to:

\[
 \lim_{P_0 \to 0}x_a =  \frac{(1-P_1) - (1-P_1) + \frac{P_0P_2}{(1-P_1)}}{P_0} =  \frac{P_2}{(1-P_1)}
\]

Simpler, we can take the condition \ref{evap2=0} and set $P_0=0$ :

\[
  P_2 + x_aP_1  -  x_a = 0
\]

Leading by chance to the same formula:

\[
  x_a = \frac{P_2}{1-P_1}
\]

A priori, if the probability of having dimers is such that we use these formulas, the probability of having empty sites will certainly not be equal to 0.

\subsection{Summary of the evaporation}

The probabilities after evaporation of the dimers are summarized by the following formulas:\\

\color{blue}
\begin{equation}
  \begin{aligned}
  \underline{P}_0 &= P_0 e^{-x}\\~\\
  \underline{P}_1 &= 1 - [1 - P_1 - x_aP_0] e^{-x_a} \\~\\
  \underline{P}_{n\ge2} &= \left[ \sum_{i=0}^n P_{n-i}\frac{x_a^i}{i!} {~}{~}
    - {~}{~} \frac{x_a^{(n-1)}}{(n-1)!} \right]  e^{-x_a}\\~\\
  ~\\
  with~&\\ ~\\
  x_a &=  \frac{(1-P_1) - \sqrt{ (1-P_1)^2 - 2P_0P_2} }{P_0} ~ if ~ P_0 \ne 0 \\
  x_a &= \frac{P_2}{1-P_1} ~ if ~ P_0 = 0
  \end{aligned}
\label{evaPn}
\end{equation}
\color{black}

\section{Appendix 4 - Proofs}
\label{annexe4}

Proofs of the validity of various tricky formulas.

\subsection{Validation of the formula (\ref{eP2}) for $P_2(x)$}
\label{peP2}

We have obtained:

\[
  P_2(x)=\frac{A+P_2}{A^2}(A-x)^2-(A-x)
  \]
  
and it is good to check that it fulfills the differential equation (\ref{equadiff_eP2A})

\[
\begin{aligned}
  P_2'(x)&=\frac{A+P_2}{A^2}[-2(A-x)]+1\\
  &=-2\frac{A+P_2}{A^2}(A-x) + 2 -1 \\
  &=\frac{1}{A-x} \left [-2\frac{A+P_2}{A^2}(A-x)^2+2(A-x) \right ] -1 \\
  &=-\frac{2}{A-x} \left [\frac{A+P_2}{A^2}(A-x)^2-(A-x) \right ]  -1   \\
  &=-\frac{2}{A-x}P_2(x)  -1   \\
\end{aligned}
\]
That establish the validity of the differential equation.

\subsection{Verification of the formula (\ref{eP3}) for $P_3(x)$}
\label{epP3}
Let's start by expanding $P_3$ to separate the different forms of contributions of $x$:
\fontsize{9}{10}\selectfont
\[
\begin{aligned}
  P_3(x) &= \left [ \frac{ P_3 + 2A}{A^2} + 2 \frac{A + P_2}{A^2}({~}ln(A)- ln(A - x){~})  \right](A-x)^2
  - 2(A - x) \\~\\
 &=  \left [\frac{ P_3 + 2A}{A^2}
  + 2\frac{(A + P_2)}{A^2}ln(A) -2 \frac{A + P_2}{A^2} ln(A - x)  \right](A-x)^2  - 2(A - x) \\~\\
&=  \left [\frac{ P_3 + 2A}{A^2}
  + 2\frac{(A + P_2)}{A^2}ln(A)\right](A-x)^2 -2 \frac{A + P_2}{A^2}ln(A - x)(A-x)^2    - 2(A - x) \\~\\
\end{aligned}
\]
\fontsize{12}{14}\selectfont
The following derivation and reorganization steps:
\fontsize{9}{10}\selectfont
\[
\begin{aligned}
  P_3'(x) &=  -2 \left [\frac{ P_3 + 2A}{A^2}  + 2\frac{(A + P_2)}{A^2}ln(A)\right](A-x)
  -2 \frac{A + P_2}{A^2} \left[ -\frac{(A - x)^2}{(A-x)} - 2ln(A - x)(A-x) \right] + 2 \\~\\
 &=  -2 \left [\frac{ P_3 + 2A}{A^2}  + 2\frac{(A + P_2)}{A^2}ln(A)\right](A-x)
  + 2 \frac{A + P_2}{A^2} \left[ (A - x) + 2ln(A - x)(A-x) \right] + 2 \\~\\
\end{aligned}
\]

\[
\begin{aligned}
&=  -2 \left [\frac{ P_3 + 2A}{A^2}  + 2\frac{(A + P_2)}{A^2}ln(A)\right](A-x)
  + 2 \frac{A + P_2}{A^2}(A - x)  + 4\frac{A + P_2}{A^2}ln(A - x)(A-x) + 2 \\~\\
  &= 2 \frac{A + P_2}{A^2}(A - x)  - 2
  -2 \left [\frac{ P_3 + 2A}{A^2}  + 2\frac{(A + P_2)}{A^2}ln(A)\right](A-x)
  + 4\frac{A + P_2}{A^2}ln(A - x)(A-x) +2 +2 \\~\\
&= \frac{2}{A-x} \left[ \frac{A + P_2}{A^2}(A - x)^2  - (A - x) \right]
  -2 \left [\frac{ P_3 + 2A}{A^2}  + 2\frac{(A + P_2)}{A^2}ln(A)\right](A-x)
  + 4\frac{A + P_2}{A^2}ln(A - x)(A-x) +4 \\~\\
&= \frac{2}{A-x}P_2(x)
  -2 \left [\frac{ P_3 + 2A}{A^2}  + 2\frac{(A + P_2)}{A^2}ln(A)\right](A-x)
  + 4\frac{A + P_2}{A^2}ln(A - x)(A-x) + 4 \\~\\
  &= \frac{2}{A-x}P_2(x) -2 \left [\frac{ P_3 + 2A}{A^2}  + 2\frac{(A + P_2)}{A^2}ln(A)
    - 2\frac{A + P_2}{A^2}ln(A - x)\right](A-x) +4 \\~\\
  &= \frac{2}{A-x}P_2(x) -2 \left \{ \left [\frac{ P_3 + 2A}{A^2}
    + 2\frac{(A + P_2)}{A^2}[{~}ln(A) - ln(A - x){~}]\right](A-x) - 2 \right \} \\~\\
&= \frac{2}{A-x}P_2(x) -\frac{2}{A-x} \left \{ \left [\frac{ P_3 + 2A}{A^2}
    + 2\frac{(A + P_2)}{A^2}[{~}ln(A) - ln(A - x){~}]\right](A-x)^2 - 2(A-x) \right \}  \\~\\
&= \frac{2}{A-x}P_2(x) -\frac{2}{A-x}P_3(x)
\end{aligned}
\]
\fontsize{12}{14}\selectfont

definitively demonstrate the validity of the formula (\ref{eP3}).

\subsection{Validation of the formula  (\ref{formulegenerale}) for $P_{n\ge3}(x)$}
\label{pePn}

Let's start with the generic formula for $f_n(x)$ and let's derivate it to check that we indeed obtain $\frac{2}{A-x}f_{n-1}(x)$.
As the generic formula (\ref{edifgenerique+}) shows us:

\fontsize{9}{10}\selectfont
\[
\begin{aligned}
  f_{(n\ge3)}(x) &= K_n + \left[ \sum_{i=1}^{n-2} K_{n-i} \frac{(-2)^i}{i!}[~ln(A-x)~]^i \right] - \frac{2^{(n-2)}}{A-x} \\~\\
  f'_{(n\ge3)}(x) &=\left[ \sum_{i=1}^{n-2} K_{n-i} \frac{(-2)^i}{i!}~(-i)\frac{[ln(A-x)~]^{i-1}}{A-x} \right]
  - \frac{2^{(n-2)}}{(A-x)^2} \\~\\
   &=\left[ \sum_{i=1}^{n-2} K_{n-i} \frac{(-2)(-2)^{i-1}}{i!}~(-i)\frac{[ln(A-x)~]^{i-1}}{A-x} \right]
  - \frac{2}{A-x}\left[ \frac{2^{(n-2-1)}}{(A-x)} \right] \\~\\
 &=\left[ \sum_{i=1}^{n-2} K_{n-i} \frac{(2)(-2)^{i-1}}{i!}(i)~\frac{[ln(A-x)~]^{i-1}}{A-x} \right]
  + \frac{2}{A-x}\left[ -\frac{2^{(n-2-1)}}{(A-x)} \right] \\~\\
  &=\left[ \sum_{i=1}^{n-2} K_{n-i} \frac{2}{A-x}\frac{(-2)^{i-1}}{(i-1)!}~[ln(A-x)~]^{i-1} \right]
  + \frac{2}{A-x}\left[ -\frac{2^{(n-2-1)}}{(A-x)} \right] \\~\\
 &=\frac{2}{A-x} \left[ \sum_{i=1}^{n-2} K_{n-i} \frac{(-2)^{i-1}}{(i-1)!}~[ln(A-x)~]^{i-1} \right]
  + \frac{2}{A-x}\left[ -\frac{2^{(n-2-1)}}{(A-x)} \right] \\~\\
\end{aligned}
\]

\[
\begin{aligned}
 &=\frac{2}{A-x} \left[ \sum_{i=0}^{n-3} K_{n-1-i} \frac{(-2)^{i}}{i!}~[ln(A-x)~]^{i} \right]
  + \frac{2}{A-x}\left[ -\frac{2^{(n-1-2)}}{(A-x)} \right] \\~\\
  &=\frac{2}{A-x} \left[  K_{n-1-0} \frac{(-2)^{0}}{0!}~[ln(A-x)~]^{0}
    + \sum_{i=1}^{n-1-2} K_{n-1-i} \frac{(-2)^{i}}{i!}~[ln(A-x)~]^{i} \right]
    + \frac{2}{A-x}\left[ -\frac{2^{(n-1)-2}}{(A-x)} \right] \\~\\
  &=\frac{2}{A-x} \left[  K_{n-1} + \sum_{i=1}^{(n-1)-2} K_{(n-1)-i} \frac{(-2)^{i}}{i!}~[ln(A-x)~]^{i} \right]
  + \frac{2}{A-x}\left[ -\frac{2^{(n-1)-2}}{(A-x)} \right] \\~\\
  &=\frac{2}{A-x} f_{n-1}(x)
\end{aligned}
\]
\fontsize{12}{14}\selectfont

The recurrence relation is verified.\\

We have furthermore:

\[
\begin{aligned}
  f_{3}(x) &= K_3 + \left[ \sum_{i=1}^{3-2} K_{3-i} \frac{(-2)^i}{i!}[~ln(A-x)~]^i \right] - \frac{2^{(3-2)}}{A-x} \\~\\
&= K_3 + \left[ K_{3-1} \frac{(-2)^1}{1!}[~ln(A-x)~]^1 \right] - \frac{2^{1}}{A-x} \\~\\
&= K_3 + \left[ K_2 (-2)[~ln(A-x)~] \right] - \frac{2}{A-x} \\~\\
&= K_3 - 2 K_2 ln(A-x)~ - ~\frac{2}{A-x} \\~\\
\end{aligned}
\]

which gives for $P_3$:

\[
P_3(x) = f_3(x)(A-x)^2  = K_3(A-x)^2 - 2 K_2 ln(A-x)(A-x)^2~ - 2(A-x)
\]

This is consistent with the formula for $P_3$ (\ref{eP3AK}) obtained earlier with \\
~\\
\[
 \begin{aligned}
  K_2 &= \frac{A+P_2}{A^2}~~~~~~~~
  and~~~~~~~~
  K_3 &= \frac{P_3 + 2A}{A^2} +  2 \frac{(A + P_2)}{A^2}ln(A)
\end{aligned}
 \]
 
 The formula (\ref{formulegenerale}) is then fully prooved by induction.

\newpage

\section{Appendix 5 - Useful computer programs}

Here, to assist in the calculation of the final size distributions for the different cases, I include several programs written in python. As the programs are in python, you will have to respect the indentation to make them work. To limit the risk of error, I will give usage examples and the obtained outputs.

\subsection{Monomer diffusion calculation}

The program below allows to calculate the statistics after the diffusion of the monomers.\\

\underline{Usage and output examples}:
\fontsize{9}{10}\selectfont
\begin{verbatim}
----------------------------------------------------------------------------
Command line :

./mdiff.py 0.4 0.3 0.2 0.1 0 0 0 0


Output :

There are  8 probabilities.
The process stops after  0.5  movements.
P 0  :  0.4  ->  0.63608160417242
P 1  :  0.3  ->  0.0
P 2  :  0.2  ->  0.16679593142097418
P 3  :  0.1  ->  0.13646939843534253
P 4  :  0.0  ->  0.04833291194585049
P 5  :  0.0  ->  0.010487925990864287
P 6  :  0.0  ->  0.0016189945994933574
P 7  :  0.0  ->  0.0001929254888594872
P 8  :  -2.7755575615628914e-17  ->  1.8674378764464476e-05
P 9  :  0.0  ->  1.519962022357588e-06
P 10  :  0.0  ->  1.0665197437654128e-07

Evolution of occupied sites :  0.6  ->  0.36391839582758

Sum of probabilities :  0.9999999930465655
----------------------------------------------------------------------------
Command line :

./mdiff.py 0.4 0.3 0.2 


Output :

There are  3 probabilities.
The process stops after  0.5  movements.
P 0  :  0.4  ->  0.63608160417242
P 1  :  0.3  ->  0.0
P 2  :  0.2  ->  0.16679593142097418
P 3  :  0.09999999999999998  ->  0.1364693984353425
P 4  :  0.0  ->  0.048332911945850474
P 5  :  0.0  ->  0.010487925990864285

Evolution of occupied sites :  0.6  ->  0.36391839582758

Sum of probabilities :  0.9981677719654513
----------------------------------------------------------------------------
Command line :

./mdiff.py 0.4 0.3  -line


Output :

0.63608160417242  0.0  0.2274489973922375  0.10614286544971084  0.02558801220662672  
 ----------------------------------------------------------------------------

\end{verbatim}
\fontsize{12}{14}\selectfont

If we give N probabilities, the program calculates the N+1th probability to make the sum equal to 1. In output we will have n+3 values.\\
The probabilities must be given in order, starting with $P_0$. \\
The program sums the calculated probabilities to check that we obtain 1 (or close to 1 if we did not provide enough probabilities). By adding the "-line" option, the output of the program is simplified: it is simply the list of the final probabilities on a line, in order to be able to easily launch the calculation of the dimer diffusion for which the program is provided below.\\

\underline{mdiff.py}:

\fontsize{9}{10}\selectfont
\begin{verbatim}
________________________________________________________________________________
#!/usr/bin/python
import math
import sys

concat=0;

nbp=len(sys.argv) - 1;

p=(nbp+4)*[0.0];
jp=(nbp+4)*[0.0];
reste=1.0;

for i in range(0,nbp) :
   if str(sys.argv[i+1]) == "-line" :
      concat = 1;
      i=i+1;
   else :
      index=i+1;
      p[i-concat]=float(sys.argv[index]);
      reste=reste - p[i-concat];

nbp = nbp-concat ;
if concat == 0 :
   print ("There are ", nbp, "probabilities.");
p[nbp]=reste;

   
xa=p[1] / (1 - p[0]);
if concat == 0 :
   print ("The process stops after ",xa," movements.");

jp[0]=(p[0]-1)*math.exp(-xa)  + 1;
for s in range(1,nbp+3) :
   jp[s]=0.0
   for i in range(0,s) : # en fait la dernière boucle c'est s-1, c'est ce qu'on veut !
       jp[s]=jp[s]+xa**i*p[s-i]/math.factorial(i);
   jp[s]=jp[s]+1/math.factorial(s)*(p[0] - 1)*(xa**s);
   jp[s]=jp[s]*math.exp(-xa);

somme=0.0;
for i in range(0,nbp+3) :
   somme=somme+jp[i];
   if concat == 0 :
      print("P",i," : ",p[i]," -> ",jp[i]);
   else :
      print(jp[i]," ",end="");

if concat == 1 :
   print("")


occupej=1-jp[0];
occupe=1-p[0];

if concat == 0 :
   print("")
   print("Evolution of occupied sites : ",occupe," -> ",occupej);
   print("");
   print("Sum of probabilities : ",somme)
   print("")
--------------------------------------------------------------------------------
\end{verbatim}
\fontsize{12}{14}\selectfont

\subsection{Dimer diffusion calculation}

The input parameters for the following programs are the same as for the program calculating the diffusion of the monomers, i.e. the list in order of the initial probabilities, starting with $P_0$. The "-line" option does not exist here, as I will not study the trimer "diffusion".\\

\underline{Usage and output example}:
\fontsize{9}{10}\selectfont
\begin{verbatim}
----------------------------------------------------------------------------
Command line :

./ddiff.py 0.5 0 0.4 0.1  0


Output :

There are  5 probabilities.

The process stops after  0.8  movements.

P 0  :  0.5  ->  0.7753355179413892
P 1  :  0.0  ->  0.0
P 2  :  0.4  ->  0.0
P 3  :  0.1  ->  0.044932896411722156
P 4  :  0.0  ->  0.07189263425875546
P 5  :  -2.7755575615628914e-17  ->  0.03594631712937772
P 6  :  0.0  ->  0.03834273827133625
P 7  :  0.0  ->  0.014378526851751084

Evolution of occupied sites :  0.5  ->  0.22466448205861078

Sum of probabilities :  0.9808286308643319
----------------------------------------------------------------------------
\end{verbatim}
\fontsize{12}{14}\selectfont

\underline{ddiff.py}:
\fontsize{9}{10}\selectfont
\begin{verbatim}
________________________________________________________________________________
#!/usr/bin/python
# calcule les nouvesses probabilités en cas de diffusion des dimères en "bloc"
import math
import sys

nbp=len(sys.argv) - 1; #Le nombre de probabilités

p=(nbp+4)*[0.0]; # les probas initiales
jp=(nbp+4)*[0.0]; # les probas finales
p[nbp]=1.0;


print ("There are ", nbp, "probabilities.")
for i in range(0,nbp) :
   p[i]=float(sys.argv[i+1]) # on assigne les probabiltés données dans le tableau
   p[nbp]=p[nbp] - p[i]; # on fait en sorte que la somme des probas soit 1

   
xa=p[2] / (1 - p[0]); 
print("");
print ("The process stops after ",xa," movements.");

jp[0]=1 - (1 - p[0])*math.exp(-xa);

s=1;
while s <= nbp+2:
   #--------------------s impair----------------------------
   jp[s]=0.0;   
   k=0
   while k <= (s-1)//2 :
       jp[s]=jp[s]+xa**k*p[s-2*k]/math.factorial(k);
       k=k+1;
   jp[s]=jp[s]*math.exp(-xa);
   #--------------------s impair----------------------------
   
   s=s+1;
   #---------------------s pair-----------------------------
   jp[s]=0.0;
   k=0
   while k <= s//2-1 :
       jp[s]=jp[s]+xa**k*p[s-2*k]/math.factorial(k);
       k=k+1;
   jp[s]=jp[s]+1/math.factorial(s//2)*(p[0] - 1)*(xa**(s/2));
   jp[s]=jp[s]*math.exp(-xa);
   #---------------------s pair-----------------------------
   s=s+1
print("")
somme=0.0;
for i in range(0,nbp+3) :
   somme=somme+jp[i];
   print("P",i," : ",p[i]," -> ",jp[i]);

occupej=1-jp[0];
occupe=1-p[0];

print("")

print("Evolution of occupied sites : ",occupe," -> ",occupej);
print("");
print("Sum of probabilities : ",somme)
print("")
--------------------------------------------------------------------------------
\end{verbatim}
\fontsize{12}{14}\selectfont

\subsection{Dimer explosion calculation}
The input parameters for the following programs are the same as for the program calculating the diffusion of the monomers, i.e. the list in order of the initial probabilities, starting with $P_0$. The "-line" option does not exist here.\\

\underline{Usage and output example}:
\fontsize{9}{10}\selectfont
\begin{verbatim}
----------------------------------------------------------------------------
Command line :

./dexplos.py 0.5 0 0.4 0.1  0


Output :

There are  5 probabilities.

The process stops after  0.2222222222222222  movements.

P 0  :  0.5  ->  0.7222222222222222
P 1  :  0.0  ->  0.0
P 2  :  0.4  ->  0.0
P 3  :  0.1  ->  0.11049876445179463
P 4  :  0.0  ->  0.09722788795395765
P 5  :  -2.7755575615628914e-17  ->  0.047815786515095215
P 6  :  0.0  ->  0.016583484825492434
P 7  :  0.0  ->  0.0044574530442513804

Evolution of occupied sites :  0.5  ->  0.2777777777777778
Sum of probabilities :  0.9988055990128135
----------------------------------------------------------------------------
\end{verbatim}
\fontsize{12}{14}\selectfont

\underline{dexplos.py}:
\fontsize{9}{10}\selectfont
\begin{verbatim}
________________________________________________________________________________
#!/usr/bin/python
# calcule les nouvesses probabilités en cas d'explosion des dimères
import math
import sys

nbp=len(sys.argv) - 1; #Le nombre de probabilités

p=(nbp+4)*[0.0]; # les probas initiales
jp=(nbp+4)*[0.0]; # les probas finales
jK=(nbp+4)*[0.0]; # les K_n 
p[nbp]=1.0;


print ("There are ", nbp, "probabilities.")
for i in range(0,nbp) :
   p[i]=float(sys.argv[i+1]) # on assigne les probabiltés données dans le tableau
   p[nbp]=p[nbp] - p[i]; # on fait en sorte que la somme des probas soit 1

A=1-p[0]; 
xa=A - A**2/(A+p[2]);

print("");
print ("The process stops after ",xa," movements.");

jp[0]=p[0]+xa;
jp[1]=0.0;
jp[2]=0.0;

jK[2] = (A + p[2])/A**2;
LNA=-2*math.log(A);
#-------------------------Calcul des constantes-----------------------
s=3;
while s <= nbp+2:
   jK[s]=( p[s] + 2**(s-2)*A ) / A**2;
   i=1
   while i <= s-2:
      jK[s] = jK[s] - jK[s-i]*(LNA**i)/math.factorial(i);
      i=i+1;
   s=s+1
   
#-------------------------Calcul des constantes-----------------------

LNA=-2*math.log(A-xa);
#-------------------------Calcul des probas-----------------------
s=3;
while s <= nbp+2:
   jp[s] = jK[s];
   i=1
   while i <= s-2:
      jp[s] = jp[s] + jK[s-i]*(LNA**i)/math.factorial(i);
      i=i+1;
   jp[s] = jp[s]*((A-xa)**2) - (2**(s-2))*(A-xa);
   s=s+1;
#-------------------------Calcul des probas-----------------------

print("")
somme=0.0
for i in range(0,nbp+3) :
   print("P",i," : ",p[i]," -> ",jp[i]);
   somme=somme+jp[i];

occupej=1-jp[0];
occupe=1-p[0];

print("")   

print("Evolution of occupied sites : ",occupe," -> ",occupej);

print("Sum of probabilities : ",somme)   
print("");
--------------------------------------------------------------------------------
\end{verbatim}
\fontsize{12}{14}\selectfont

\newpage
\subsection{Dimer evaporation calculation}

The input parameters for the following programs are the same as for the program calculating the diffusion of the monomers, i.e. the list in order of the initial probabilities, starting with $P_0$. The "-line" option does not exist here.\\

\underline{Usage and output example}:
\fontsize{9}{10}\selectfont
\begin{verbatim}
----------------------------------------------------------------------------

Command line :

./devap.py 0.5 0 0.4 0.1  0


Output :

There are  5 probabilities.
The process stops after  0.4508066615170332  movements.

P 0  :  0.5  ->  0.3185570044635628
P 1  :  0.0  ->  0.5064936107579598
P 2  :  0.4  ->  3.5366932101106616e-17
P 3  :  0.1  ->  0.11872237419187774
P 4  :  0.0  ->  0.04543713198682613
P 5  :  -2.7755575615628914e-17  ->  0.009318277295881153
P 6  :  0.0  ->  0.0013162491387744963
P 7  :  0.0  ->  0.00014199246540335273

Evolution of occupied sites :   0.5  ->  0.6814429955364372

Sum of probabilities :  0.9999866403002855

----------------------------------------------------------------------------
\end{verbatim}
\fontsize{12}{14}\selectfont

\underline{devap.py}:
\fontsize{9}{10}\selectfont
\begin{verbatim}
________________________________________________________________________________
#!/usr/bin/python
# calcule les nouvesses probabilités en cas d'évaporation
import math
import sys

nbp=len(sys.argv) - 1; #Le nombre de probabilités

p=(nbp+4)*[0.0]; # les probas initiales
jp=(nbp+4)*[0.0]; # les probas finales
p[nbp]=1.0;

print ("There are ", nbp, "probabilities.")
for i in range(0,nbp) :
   p[i]=float(sys.argv[i+1]) # on assigne les probabiltés données dans le tableau
   p[nbp]=p[nbp] - p[i]; # on fait en sorte que la somme des probas soit 1

a=(1-p[1]);
b=2*p[0]*p[2];

if p[0] == 0 :
   xa=p[2] / a ;
else :
   xa=(a - (a*a - b)**(1/2))/p[0];

print ("The process stops after ",xa," movements.");

jp[0]=p[0]*math.exp(-xa);
jp[1]=1-(1-p[1]-xa*p[0])*math.exp(-xa);

s=2;
while s <= nbp+2:
   jp[s]=0.0;
   
   k=0
   while k <= s :
       jp[s]=jp[s]+p[s-k] * (xa**k) / math.factorial(k);
       k=k+1;
   jp[s]=jp[s] - (xa**(s-1)) / math.factorial(s-1);
   jp[s]=jp[s]*math.exp(-xa);
   s=s+1;

somme=0.0;
print("")   
for i in range(0,nbp+3) :
   somme=somme+jp[i];
   print("P",i," : ",p[i]," -> ",jp[i]);

occupej=1-jp[0];
occupe=1-p[0];

print("")   

print("Evolution of occupied sites :  ",occupe," -> ",occupej);
print("")   
print("Sum of probabilities : ",somme)   
print("")   
--------------------------------------------------------------------------------
\end{verbatim}
\fontsize{12}{14}\selectfont

\end{document}